\documentclass[aps,prl,twocolumn,reprint,superscriptaddress,floatfix]{revtex4-1} 
\usepackage[T1]{fontenc}
\usepackage[utf8]{inputenc}
\usepackage{newtxtext}
\usepackage{newtxmath}
\usepackage{amsmath}
\usepackage{graphicx}   
\usepackage{siunitx}    
\usepackage[english]{babel}
\usepackage[version=4]{mhchem}
\usepackage{microtype}
\usepackage{caption}
\usepackage{subcaption}

\usepackage{wasysym}

\usepackage{verbatim}

\usepackage[colorlinks=true,allcolors=blue,unicode]{hyperref}

\newcommand{\bnmr}{$\beta$-NMR}

\newcommand{\eli}{\textsuperscript{8}{Li}}
\newcommand{\elip}{\textsuperscript{8}{Li}\textsuperscript{+}}

\newcommand{\lip}{{Li}\textsuperscript{+}}

\newcommand{\tio}{\ce{TiO2}}

\begin{document}

\title{Bi-Arrhenius diffusion and surface trapping of $^{8}$Li$^{+}$ in rutile TiO$_2$}


\author{A. Chatzichristos}
\email{aris.chatzichristos@alumni.ubc.ca}
\affiliation{Department of Physics and Astronomy, University of British Columbia, Vancouver, BC V6T~1Z1, Canada}
\affiliation{Stewart Blusson Quantum Matter Institute, University of British Columbia, Vancouver, BC V6T~1Z4, Canada}

\author{R. M. L. McFadden}
\affiliation{Stewart Blusson Quantum Matter Institute, University of British Columbia, Vancouver, BC V6T~1Z4, Canada}
\affiliation{Department of Chemistry, University of British Columbia, Vancouver, BC V6T~1Z4, Canada}

\author{M. H. Dehn}
\affiliation{Department of Physics and Astronomy, University of British Columbia, Vancouver, BC V6T~1Z1, Canada}
\affiliation{Stewart Blusson Quantum Matter Institute, University of British Columbia, Vancouver, BC V6T~1Z4, Canada}

\author{S. R. Dunsiger}
\affiliation{Department of Physics, Simon Fraser University, Burnaby, BC V5A~1S6, Canada}	

\author{D. Fujimoto}
\affiliation{Department of Physics and Astronomy, University of British Columbia, Vancouver, BC V6T~1Z1, Canada}
\affiliation{Stewart Blusson Quantum Matter Institute, University of British Columbia, Vancouver, BC V6T~1Z4, Canada}

\author{V. L. Karner}
\affiliation{Stewart Blusson Quantum Matter Institute, University of British Columbia, Vancouver, BC V6T~1Z4, Canada}
\affiliation{Department of Chemistry, University of British Columbia, Vancouver, BC V6T~1Z4, Canada}


\author{I. McKenzie}
\affiliation{TRIUMF, 4004 Wesbrook Mall, Vancouver, BC V6T~2A3, Canada}	
\affiliation{Department of Chemistry, Simon Fraser University, Burnaby, BC V5A~1S6, Canada}	

\author{G. D. Morris}
\affiliation{TRIUMF, 4004 Wesbrook Mall, Vancouver, BC V6T~2A3, Canada}

\author{M. R. Pearson}
\affiliation{TRIUMF, 4004 Wesbrook Mall, Vancouver, BC V6T~2A3, Canada}

\author{M. Stachura}
\affiliation{TRIUMF, 4004 Wesbrook Mall, Vancouver, BC V6T~2A3, Canada}

\author{J. Sugiyama}
\affiliation{Toyota Central Research and Development Laboratories, Inc., Nagakute, Aichi 480-1192, Japan}
\affiliation{Advanced Science Research Center, Japan Atomic Energy Agency, Tokai, Ibaraki 319-1195, Japan}

\author{J. O. Ticknor}
\affiliation{Stewart Blusson Quantum Matter Institute, University of British Columbia, Vancouver, BC V6T~1Z4, Canada}
\affiliation{Department of Chemistry, University of British Columbia, Vancouver, BC V6T~1Z4, Canada}

\author{W. A. MacFarlane}
\affiliation{Stewart Blusson Quantum Matter Institute, University of British Columbia, Vancouver, BC V6T~1Z4, Canada}
\affiliation{Department of Chemistry, University of British Columbia, Vancouver, BC V6T~1Z4, Canada}
\affiliation{TRIUMF, 4004 Wesbrook Mall, Vancouver, BC V6T~2A3, Canada}

\author{R. F. Kiefl}
\email{kiefl@triumf.ca}
\affiliation{Department of Physics and Astronomy, University of British Columbia, Vancouver, BC V6T~1Z1, Canada}
\affiliation{Stewart Blusson Quantum Matter Institute, University of British Columbia, Vancouver, BC V6T~1Z4, Canada}
\affiliation{TRIUMF, 4004 Wesbrook Mall, Vancouver, BC V6T~2A3, Canada}

\date{\today}

\begin{abstract}
We report measurements of the diffusion rate of isolated ion-implanted \elip\ within \SI{\sim 120}{\nano\meter} of the surface of oriented single-crystal rutile \ce{TiO2} using a radiotracer technique. 
The \mbox{$\alpha$-particles} from the \eli\ decay provide a sensitive monitor of the distance from the surface and how the depth profile of \eli\ evolves with time. 
The main findings are that the implanted \ce{Li^{+}} diffuses and traps at the $(001)$ surface.
The $T$-dependence of the diffusivity is described by a bi-Arrhenius expression with activation energies of \SI{0.3341 \pm  0.0021}{\electronvolt} above \SI{200}{\kelvin}, whereas at lower temperatures it has a much smaller barrier of \SI{0.0313 \pm 0.0015}{\electronvolt}. 
We consider possible origins for the surface trapping, as well the nature of the low-$T$ barrier.
\end{abstract}

\maketitle
It is well known~\cite{1964-Johnson-PR-136-A284,1966-Johnson-JAP-37-668} that \ce{Li^{+}} diffusion in rutile \tio\ through the $c$-axis channels is extremely fast, greatly surpassing all other interstitial cations~\cite{2010-VanOrman-RMG-72-757}, with a room temperature diffusion coefficient exceeding many modern solid-state \ce{Li} electrolytes~\cite{2016-Bachman-CR-116-140}.
A major limitation to its use as an electrode material in Li-ion batteries is its limited \ce{Li} uptake at room temperature~\cite{MURPHY1983413, ZACHAUCHRISTIANSEN19881176};
however, the discovery that using nanosized crystallites mitigates this issue~\cite{2006-Hu-AM-18-1521} has led to renewed interest in its applicability~\cite{2013-Reddy-CR-113-5364}.

There are a number of poorly understood aspects of rutile lithiation, including the cause of the limited \ce{Li^{+}} uptake, as well as why reported Li diffusion rates differ by orders of magnitude, even under the same experimental conditions~\cite{1964-Johnson-PR-136-A284, 1987-Kanamura-JPS-20-127, 2004-Churikov-RJE-40-63, 2010-Bach-EA-55-4952, 2014-Churikov-JSSE-18-1425, 2017-McFadden-CM-29-10187}.
Theoretical studies (see e.g.,~\cite{2002-Koudriachova-PRB-65-235423, 2006-Gligor-SSI-177-2741, 2009-Kerisit-JPCC-113-20998, 2012-Yildirim-PCCP-14-4565, 2014-Kerisit-JPCC-118-24231, 2014-Jung-AIPA-4-017104, 2015-Arrouvel-CTC-1072-43}) have been unable to reproduce the characteristics of \ce{Li^{+}} migration found in experiments~\cite{1964-Johnson-PR-136-A284, 2010-Bach-EA-55-4952, 2017-McFadden-CM-29-10187}. 
A direct technique applicable to the nanoscale could help resolve these issues. 
To this end, we developed a variation to the classical radiotracer method, the \eli\ \mbox{$\alpha$-radiotracer} method, which uses the attenuation of the progeny \mbox{$\alpha$-particles} from the radioactive decay of \eli, to study nanoscale \ce{Li} diffusion. 
This method differs from conventional radiotracer diffusion experiments in several key aspects: (a) it is non-destructive (b) it is sensitive to motions on the nanometer scale~\cite{2016-Ishiyama-NIMB-376-379} (c) it can be applied to thin films and heterostructures and (d) it is amenable for the use of  short-lived isotopes ($\tau_{1/2}\sim$\SI{1}{\second}). 

In this study, we employ the \mbox{$\alpha$-radiotracer} method to extract the diffusion coefficient and its activation energy for isolated \ce{Li} in rutile \ce{TiO2} and show that \lip\ traps at the $(001)$ rutile surface. 
In addition, we report that the nanoscale \ce{Li} diffusion exhibits bi-Arrhenius behavior. 
The high-$T$ (above \SI{\sim 200}{\kelvin}) activation energy and diffusion rate are in agreement with previous studies. 
The low-$T$ behavior is discussed in the context of the recently reported Li-Ti$^{3+}$ polaron complex~\cite{2017-McFadden-CM-29-10187}; here we suggest that part of that signal is also connected to \ce{Li} hopping/diffusion. 

The experiment was performed using the ISAC facility at TRIUMF~\cite{2014-Morris-HI-225-173}, in Vancouver, Canada.
The samples were commercial chemo-mechanically polished (roughness \SI{<0.5}{\nano\meter}) single crystal rutile \ce{TiO2} substrates (CRYSTAL GmbH) with typical dimensions of \SI{7 x 7 x 0.5}{\milli\meter}. 
The surfaces were free of macroscopic defects under 50x magnification.

In the experiment, a short beam pulse of low energy ($0.1-30$~\SI{}{\kilo\electronvolt}) \elip\ ions is implanted close to the surface of the rutile targets (at an average depth of $\sim$\SI{100}{\nano\meter}) housed in an ultra-high vacuum cold finger cryostat~\cite{2014-Morris-HI-225-173,2004-Salman-PRB-70-104404}. 
The energy of the beam defines the initial \lip\ implantation profile. 
Upon arrival, the \ce{^{8}Li^{+}} starts to diffuse through the sample and undergoes \mbox{$\beta$-decay} to \ce{^{8}Be} which then decays (immediately) into two energetic \mbox{$\alpha$-particles}, each with a mean energy of \SI{1.6}{\mega\electronvolt}. 
Due to their rapid attenuation inside the sample, the highest energy \mbox{$\alpha$-particles} escaping the sample originate from \elip\ that have diffused back closer to the surface. 

\begin{figure}[tb]
	\centering
	\captionsetup{justification=raggedright,singlelinecheck=false}
	\includegraphics[width=0.33\textwidth, keepaspectratio]{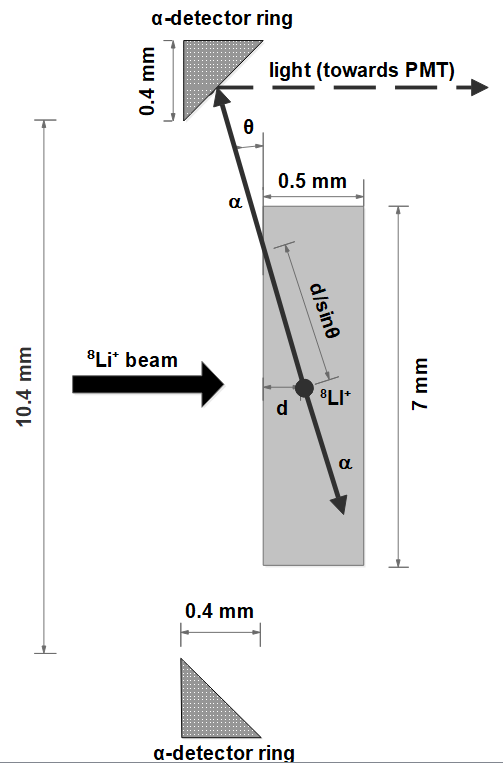}
	\caption{ \label{fig:GeometricEnhancement}
		Schematic of the ultra-high vacuum ($10^{-10}$~Torr) sample region showing the cross-section of the ring detector.
		The \mbox{$\alpha$-particles} originating at depth $d$ that reach the \mbox{$\alpha$-detector} traverse distance $d/\sin\theta$~\cite{2013-Ishiyama-JJAP-52-010205} through the sample. 
		Not to scale.
	}
\end{figure}

To further amplify the sensitivity to \ce{^{8}Li^{+}} near the surface, the \mbox{$\alpha$-detector} is placed at a grazing angle, $\theta \leq 4.4^\mathrm{o}$, relative to the surface, as shown in Fig.~\ref{fig:GeometricEnhancement}. 
The \mbox{$\alpha$-detector} in our setup is an \ce{Al} ring, whose inside surface is cut at \SI{\sim45}{\degree} and coated with a thin layer of \ce{Ag}-doped \ce{ZnS}, a well known scintillator sensitive to \mbox{$\alpha$-particles}~\cite{1959-Asada-JPSJ-14-12}. 
The light from the ZnS:Ag scintillator is collected in the forward direction using two \SI{5}{\centi\meter} $\diameter$ lenses which focus the light onto the photo-cathode of a fast photomultiplier (PMT).
The second lens and the PMT are positioned outside the vacuum chamber, behind a transparent viewport. 
The PMT pulses have a large signal to noise ratio ($>10$) and pass through a timing filter amplifier to be discriminated, so that only the top $1/3$ of pulses above the noise level are counted.

\begin{figure}[tb]
	\centering
\captionsetup{justification=raggedright,singlelinecheck=true}
	\includegraphics[width=0.4\textwidth]{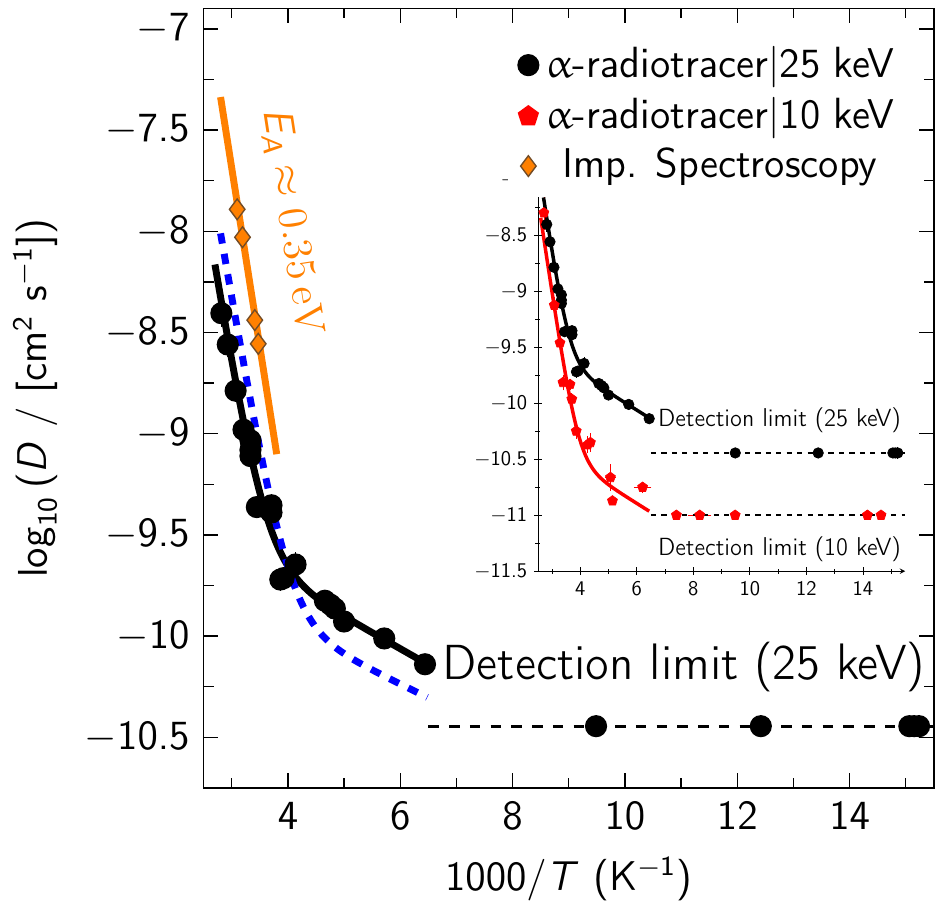}
	\caption{ \label{fig:arrhenius-all}
		Arrhenius plot, comparing reported \ce{Li} diffusivity in rutile \ce{TiO2}~\cite{2010-Bach-EA-55-4952,2017-McFadden-CM-29-10187}.
		The solid black line is the bi-Arrhenius fit of Eq.~\ref{eq:bi-Arrhenius} with $P_{tr} =$\SI{100}{\percent}.
		The blue dashed line is the fit of the present data using the two Arrhenius components found with \bnmr~\cite{2017-McFadden-CM-29-10187}, assuming that only a fraction $f$ of the \bnmr\ fluctuations corresponds to a hop. which
		$f$ was the only free parameter and yielded $f =$\SI{28.1 \pm 1.7 e-6}. 
		\textit{Insert:} \mbox{$\alpha$-radiotracer} data acquired with \elip\ beam energies of 10 and \SI{25}{\kilo\electronvolt}.
	}
\end{figure}

The diffusion rate of \ce{Li} inside the sample is directly related to the time it takes to reach the surface, which in turn relates to the \mbox{$\alpha$-rate} as a function of time. 
This method has intrinsic time- and length-scales of $\tau_{1/2}$\SI{\sim1}{\second} and $d$\SI{\sim100}{\nano\meter}, respectively, which leads to a theoretical sensitivity to the diffusion rate $D$ from $10^{-12}$ to \SI{e-8}{\centi\meter\squared\per\second}.
This technique thus covers an optimal range of $D$ for battery materials. 
However, our effective sensitivity limit is closer to \SI{e-11}{\centi\meter\squared\per\second}, determined by experimental factors such as the finite counting statistics and the existence of small distortions  due to pileup in the detector response. 
In addition, the experimental sensitivity is somewhat higher for a lower implantation energy (see insert of Fig.~\ref{fig:arrhenius-all}).  

In situations where \ce{Li^{+}} is immobile, the probability of detecting an $\alpha$ for any given decay event is time-independent and the measured \mbox{$\alpha$-counts} follow the decay rate of \eli. 
This can be monitored conveniently using the high energy \mbox{$\beta$-particles} from the \eli\ decay, which are weakly attenuated over these distances.
Thus, the ratio of counts $Y_{\alpha}=N_{\alpha}/N_{\beta}$ is constant in time. 
On the other hand, when \ce{Li^{+}} is mobile, the ratio is time-dependent when the mean diffusion length in the \eli\ lifetime is comparable to the mean depth of implantation, reflecting the fact that the \elip\ depth distribution is evolving in time.
The information on \ce{Li} diffusion comes from the time evolution of the \mbox{$\alpha$-signal}. 
The absolute \mbox{$\alpha$-to-$\beta$} ratio, i.e., the \textit{baseline} ratio of $Y_{\alpha}$, in the absence of diffusion, depends on experimental factors such as detector efficiencies, therefore in order to account for these systematics, each \mbox{$\alpha$-spectrum} is self-normalized to start from unity at time zero, i.e., $Y^n_{\alpha}(t)=Y_{\alpha}(t)/Y_{\alpha}(0)$~\cite{2015-Ishiyama-NIMB-354-297}.

\begin{figure}[tb]
	\centering
		\includegraphics[width=0.45\textwidth]{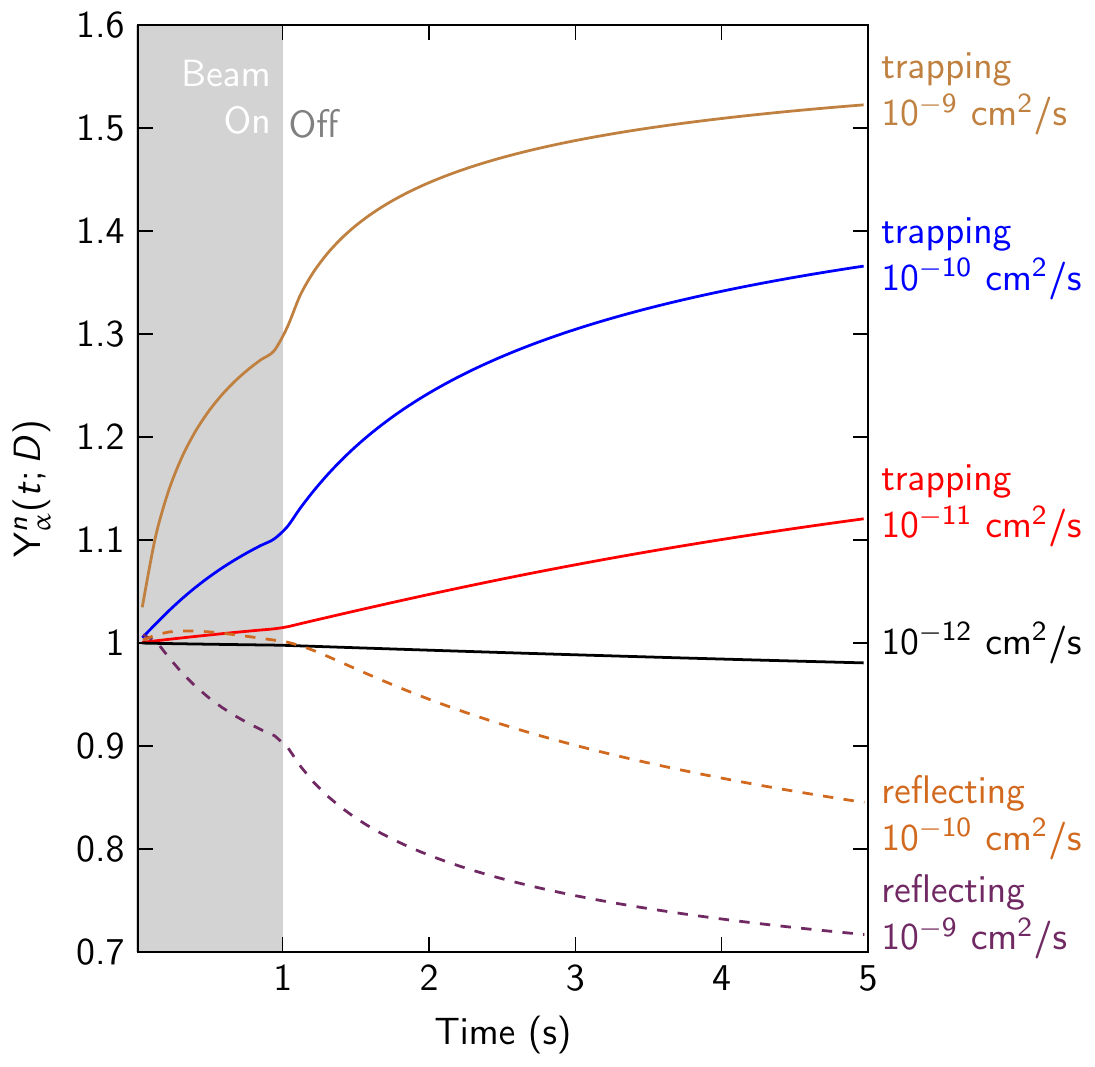}
	\captionsetup{justification=raggedright,singlelinecheck=false}
	\caption{\label{fig:simulated-signal-both}
		The calculated normalized \mbox{$\alpha$-yield}, $Y^n_{\alpha}(t;D)$, as a function of time in \ce{TiO2}, for a beam pulse of \SI{1}{\second}, different diffusion rates and an initial beam energy of \SI{25}{\kilo\electronvolt}, for a fully trapping (solid) and fully reflective (dashed) surface. 
		The time evolution of the \mbox{$\alpha$-signal} indicates both the diffusion rate and the surface boundary condition. 
		In both cases $Y^n_{\alpha}(t;D)$ is flat for $D\sim$\SI{e-12}{\centi\meter\squared\per\second} and differs from that line increasingly with increasing $D$. 
		Note that the spectra evolve towards transient equilibrium during the beam pulse, while after the implantation they evolve towards unreplenished equilibrium, resulting in a ``kink'' at \SI{1}{\second}.	
	}
\end{figure}

In order to extract the \ce{Li} diffusion rate, the experimentally acquired normalized $Y^n_{\alpha}(t)$ was compared to a library of simulated $Y^n_{\alpha}(t)$ signals. 
To this end, we performed numerical solutions to Fick's laws in 1D to generate the time-evolved depth distribution of \elip, accounting for the boundary conditions of the crystal surface and the initial \elip\ stopping profile as simulated by the SRIM Monte Carlo package~\cite{SRIM}. 
$Y^n_{\alpha}(t;D)$ is then obtained by multiplying each bin of the depth profile of \eli\ with the probability of detecting an $\alpha$ emitted at that depth.
The\mbox{ $\alpha$-detection} probability versus depth was extracted using the Geant4~\cite{Geant4-dev-2016} simulation package. 
The qualitative characteristics of $Y^n_{\alpha}(t;D)$ were found to depend heavily on whether the diffusing \lip\ ions accumulate or reflect upon reaching the sample surface (see Fig.~\ref{fig:simulated-signal-both}), implying that one can unambiguously infer the \elip\ behavior at the surface, i.e., whether it is trapped or reflected. 
Furthermore, $Y^n_{\alpha}(t;D)$ deviates significantly from the simple exponential decay of \eli\ ($\propto \exp[-t/\tau]$) with increasing diffusion rate. 

With an accumulating boundary condition at the surface, faster diffusion results in a monotonically increasing $Y^n_{\alpha}(t;D)$, while a reflecting boundary condition leads to $Y^n_{\alpha}(t;D)$ that decreases with time, since the overall mean distance from the surface will then increase with time as the \ce{Li} migrates away from the surface back to the bulk of the sample, towards the uniform depth distribution. 
Between these two ideal cases, there could be a non-zero trapping probability $P_{tr}$ at the sample's surface. 
For a fixed diffusivity $D$, $Y^n_{\alpha}(t;D,P_{tr})$ gradually evolves from looking reflective-like ($P_{tr}<$\SI{20}{\percent}), towards resembling the accumulating condition (for $P_{tr}\geq$\SI{50}{\percent}).
Note that after each reflection, the \ce{Li^{+}} will continue their random walk, so for any non-zero value of $P_{tr}$, most Li ions will eventually (after several reflections) get trapped at the surface if the diffusion is fast enough.

A technique similar to the one discussed here has been developed by Jeong et al.~\cite{2005-Jeong-NIMB-230-596} for \ce{Li^{+}} diffusion on micrometer and, recently, by Ishiyama et al.~\cite{2016-Ishiyama-NIMB-376-379} on a nanometer length scales; however, the experiment reported here differs in a few key ways. 
In particular, the \eli\ implantation rates accessible at TRIUMF (typically 10$^6$-10$^7$ \elip/s) are 1-2 orders of magnitude larger~\cite{2016-Ishiyama-NIMB-376-379}, which allows the \mbox{$\alpha$-detector} to be placed at a grazing angle $\theta$ ($\leq$\SI{4.4}{\degree} versus \num{10 \pm 1}$^\mathrm{o}$~\cite{2015-Ishiyama-NIMB-354-297}). 
This detector configuration significantly decreases the \mbox{$\alpha$-counts}, but greatly enhances the sensitivity to the near-surface region. 
In addition, the ZnS:Ag ring detector used in the present setup is much simpler and easier to install close to the sample in UHV compared to a Si detector~\cite{2015-Ishiyama-NIMB-354-297}, although it has less energy  resolution.

Using this technique, we performed \mbox{$\alpha$-radiotracer} measurements on rutile \tio\ at various temperatures with two beam energies (10 and \SI{25}{\kilo\electronvolt}) and two sample orientations. 
As \ce{Li^{+}} is known to diffuse primarily along the $c$-axis of rutile, 
if the $c$-axis is oriented parallel to the surface (perpendicular to the beam), then the \elip\ motion should not change the initial implantation profile. 
Since the $ab$-plane diffusivity $D_{ab}\ll$\SI{e-12}{\centi\meter\squared\per\second}, $Y^n_{\alpha}(t)$ is expected to be time-independent. 
On the other hand, if the $c$-axis is perpendicular to the surface, then the depth distribution of lithium should be evolving with time, since $D_c\gg$\SI{e-12}{\centi\meter\squared\per\second}.

\begin{figure}[tb]
	\centering
	\captionsetup{justification=raggedright,singlelinecheck=true}
	\includegraphics[width=0.45\textwidth]{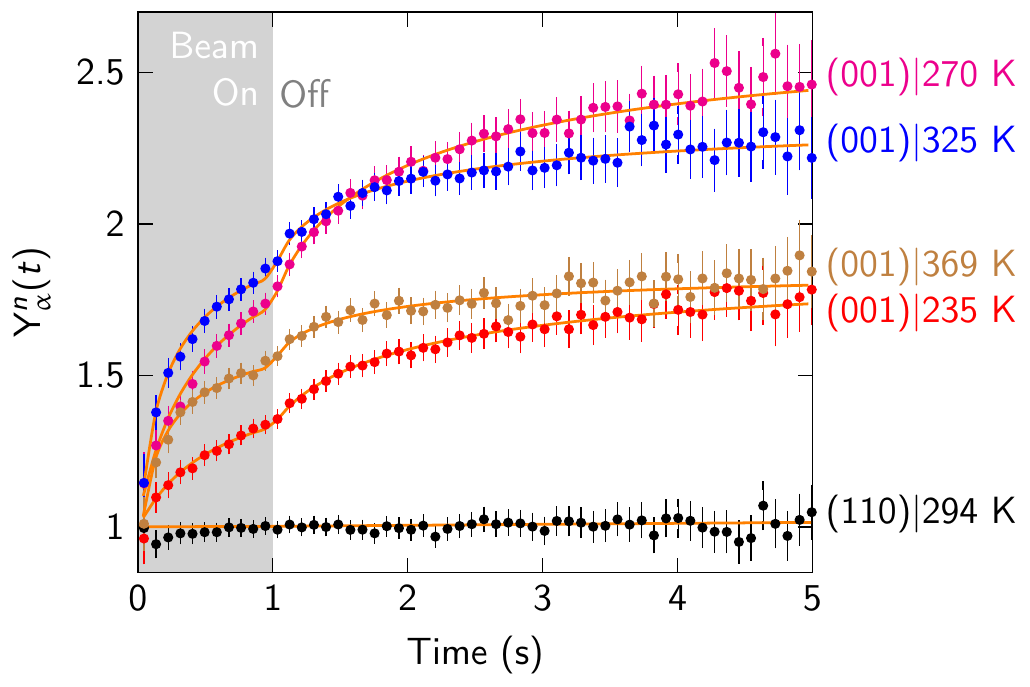}
	\caption{ \label{fig:fitted-spectrum}
		Comparison of the measured normalized \mbox{$\alpha$-yield} $Y^n_{\alpha}(t;D,P_{tr})$ for the $(110)$- and $(001)$-oriented rutile \ce{TiO2} and fits (orange lines) for a beam energy of \SI{25}{\kilo\electronvolt} and a surface trapping probability $P_{tr}=1$. 
		The increasing signal with time in the $(001)$ crystal is consistent with the anisotropy of the \ce{Li} diffusion coefficient~\cite{1964-Johnson-PR-136-A284} and indicates that \ce{Li} diffuses fast along the $c$-axis and gets trapped upon reaching the sample surface. 
		For increasing temperature, $Y^n_{\alpha}(t;D)$ saturates more rapidly, indicating that above room temperature, most of \ce{Li} gets trapped at the $(001)$-surface during its lifetime. 		
		In contrast, the diffusion coefficient along the $ab$-plane is smaller than the theoretical detection limit of \SI{\sim e-12}{\centi\meter\squared\per\second}, yielding $Y^n_{\alpha}=1$ independent of time.
		Above room temperature, the $c$-axis normalized spectra $Y^n_{\alpha}(t;T)$ get progressively suppressed, as the normalization factor $Y_{\alpha}(t=0;T)$ increases substantially due to fast diffusion.   
	}
\end{figure}
In Fig.~\ref{fig:fitted-spectrum} we compare the measured normalized \mbox{$\alpha$-yield}, $Y^n_{\alpha}$, for the $c$-axis parallel and perpendicular to the surface. 
As expected, the time spectrum for the $(110)$ orientation of \tio\ rutile ($c$-axis parallel to the surface) is completely flat at \SI{294}{\kelvin}, indicating that the $ab$-plane diffusion rate is lower than the theoretical detection limit \SI{\sim e-12}{\centi\meter\squared\per\second} (Fig.~\ref{fig:simulated-signal-both}), consistent with other studies reporting an $ab$-plane diffusion rate of \SI{e-15}{\centi\meter\squared\per\second} or lower~\cite{1964-Johnson-PR-136-A284}.
Also shown in Fig.~\ref{fig:fitted-spectrum} are a few examples of experimental data for the $(001)$ orientation in the range of \SIrange{60}{370}{\kelvin}, with the corresponding fits to the model described above.

To fit the data, we used a custom C++ code applying the MINUIT~\cite{1975-James-CPC-10-343} minimization functionalities of ROOT~\cite{1997-Brun-NIMA-389-81} to compare the $Y^n_{\alpha}$ signals to the library of calculated spectra.
The free parameters of the fit were $D$ and $P_{tr}$.
All $Y^n_{\alpha}(t;D,P_{tr})$ spectra at both implantation energies (\SI{10}{\kilo\electronvolt} and \SI{25}{\kilo\electronvolt}) were fitted simultaneously with a shared $P_{tr}$ value. 
For the $(001)$ orientation $Y^n_{\alpha}$ increases rapidly with time, approaching saturation, indicating that lithium diffuses fast along the $c$-axis and gets trapped at (or within few \si{\nano\meter} of) the surface (see Fig.~\ref{fig:simulated-signal-both}).
For $P_{tr}\geq$\SI{50}{\percent}, the global ${\chi}^2$ value is completely insensitive to $P_{tr}$, but for $P_{tr}<$\SI{50}{\percent}, the quality of the fits deteriorates rapidly. 
This is the first \textit{unambiguous} evidence for Li trapping (with at least \SI{50}{\percent} probability) at the $(001)$ surface.
There is no evidence of Li de-trapping up to \SI{370}{\kelvin}, since at that temperature $Y^n_{\alpha}(t;T)$ reaches saturation after \SI{\sim 2}{\second} and any Li surface de-trapping would lead to an observable decrease of $Y^n_{\alpha}(t;T)$ at later times. 
The non-zero trapping probability is most likely related to the reported difficulty of intercalating Li into rutile, as the Li ions would tend to stick at or near the surface rather than diffusing into the bulk. 

It is not clear whether the \ce{Li^{+}} surface trapping is caused by an electrostatic potential well (similar to \ce{H} in \ce{Pd}~\cite{1998-Okuyama-SS-401-3}), a partially reconstructed surface~\cite{2003-Diebold-SSR-48-53}, or by a chemical sink either due to an adsorbate, or a solid state reaction at the surface (e.g., forming cubic \ce{LiTiO2}). 
Subsequent measurements of an adsorbate-free rutile sample, as well as samples capped with thin layers of materials capable of altering the surface chemistry are needed to resolve this question. 
        
Turning to the values of $D(T)$ extracted using the above analysis (see Fig.~\ref{fig:arrhenius-all}), they reveal a bi-Arrhenius relationship of the form:
\begin{equation} \label{eq:bi-Arrhenius}
   D(T) = D_{H} \exp \left [ -E_{H} / (k_{B}T) \right ] + D_{L} \exp \left [ -E_{L} / (k_{B}T) \right ],
\end{equation}
where $E_{i}$ is the activation energy and $D_{i}$ is the prefactor of each component. 
These were found to be $E_H = \SI{0.3341 \pm  0.0021}{\electronvolt}$ and  $D_H = \SI{2.31 \pm 0.18 e-4}{\centi\meter\squared\per\second}$ for the high-$T$ component and $E_L = \SI{0.0313 \pm 0.0015}{\electronvolt}$ and  $D_L = \SI{7.7 \pm 0.7 e-10}{\centi\meter\squared\per\second}$ for the low-$T$ component, respectively.
This extracted $E_{H}$ is in excellent agreement with values deduced by other techniques~\cite{1964-Johnson-PR-136-A284, 2010-Bach-EA-55-4952, 2017-McFadden-CM-29-10187} and the diffusion rates at high temperatures are very similar to the ones found in rutile nanorods using impedance spectroscopy~\cite{2010-Bach-EA-55-4952}.

Both data sets acquired with beam energies of 10 and \SI{25}{\kilo\electronvolt} yield virtually the same bi-Arrhenius activation energies and they are in agreement at high-$T$, but the low-$T$ component of the \SI{10}{\kilo\electronvolt} data is shifted lower by about an order of magnitude.  
For trapping probability $P_{tr} <$\SI{100}{\percent}, the apparent gap narrows and for $P_{tr}=$\SI{50}{\percent} is about half an order of magnitude wide. 
The persistency of the gap suggests that it might be related to either a discrepancy between the SRIM and the actual implantation profiles (e.g., due to channeling~\cite{2003-Beals-PB-326-1}), or due to some small random disorder close to the surface parameterized by some energy scale ($\Delta$). 
At higher temperatures when $kT >>\Delta$ its effect would diminish. 
Both these effects would affect the (closer to the surface) \SI{10}{\kilo\electronvolt} data more than the \SI{25}{\kilo\electronvolt} and would become irrelevant at fast diffusivities, explaining the agreement of the two sets at high temperatures and why the diffusion seems slower at low-$T$ for the \SI{10}{\kilo\electronvolt} data. 

A bi-Arrhenius relationship for diffusivity is not uncommon; in vacancy ion conductors~\cite{1988-Hooton-CJC-66-4}, it may occur from a crossover between a region at high-$T$, where vacancies are thermally generated, to a region at lower $T$ with a shallower slope. 
As \mbox{$\alpha$-radiotracer} is always only measuring the diffusion of \lip, rather than the net ionic conductivity, the origin of the two Arrhenius components can't be the same as above. 
While we cannot be conclusive about it, we consider some possibilities.  
 
We first consider a recent \bnmr\ experiment on rutile~\cite{2017-McFadden-CM-29-10187}, which also used an implanted \elip\ beam on similar crystals. 
The \bnmr\ measurements revealed two peaks in the relaxation rate $1/T_1$, one below \SI{100}{\kelvin} and one above \SI{200}{\kelvin}.

Below \SI{100}{\kelvin}, a \SI{0.027}{\electronvolt} barrier was attributed to dynamics of electron-polarons in the vicinity of the implanted ion~\cite{2009-Kerisit-JPCC-113-49,2013-Brant-JAP-113-5}.
In principle, these dynamics might not be diffusive, e.g. if the \elip\ is static and the polaron is thermally trapped by the Li and cycles through trapping and detrapping. 
Nonetheless, our current measurement really shows that there is some long range diffusion of \elip\ at low-$T$, with a barrier significantly different than high-$T$. 
While our $E_L$ is of a similar magnitude to that found with \bnmr, it is also compatible with the diffusion barrier predicted from theory for isolated \ce{Li} in rutile~\cite{2001-Koudriachova-PRL-86-1275, 2002-Koudriachova-PRB-65-235423, 2006-Gligor-SSI-177-2741, 2009-Kerisit-JPCC-113-20998, 2012-Yildirim-PCCP-14-4565, 2014-Kerisit-JPCC-118-24231, 2014-Jung-AIPA-4-017104, 2015-Arrouvel-CTC-1072-43}. 
The \mbox{$\alpha$-radiotracer} cannot distinguish whether \ce{Li} moves either as a simple interstitial, or as part of a Li-polaron complex, it would only identify their weighted average contribution to the motion of \ce{^{8}Li^{+}}.
The similarity of the observed activation energy at low temperatures to the theoretical value suggests that a small fraction of the \ce{Li^{+}} interstitials does not combine with a polaron, but rather diffuses as a simple ion. 
If this fraction is small, that would explain why the low-$T$ prefactor is so much smaller than the high-$T$.  
 
It seems possible that the larger activation energy observed above  \SI{200}{\kelvin} may involve diffusion of a more complex object, possibly a Li-polaron complex, or it could be related to a disassociation energy of \lip\ with the polarons, which are known to form Coulomb bond defect complexes. 
Indeed, theory predicts a diffusion barrier of \SI{0.29}{\electronvolt} for the Li-polaron complex and a disassociation energy of \SI{0.45}{\electronvolt}~\cite{2009-Kerisit-JPCC-113-49}, both comparable to the high-$T$ barrier found here.
The Li-polaron complex is overall electrically neutral, so its movement should contribute to the diffusivity of Li but not to the ionic conductivity in terms of charge transport. 
An electric field would not cause it to move - unless it was strong enough to destabilize the complex (strong potential gradient).
Thus, if it is a neutral Li-polaron complex moving at high-$T$, one would expect the impedance measurement to yield a very different Arrhenius slope. 

The much larger prefactor above \SI{200}{\kelvin}, compared to low-$T$, is also quite intriguing and is further evidence that these are two very different mechanisms for diffusion of Li in rutile. 
Indeed, $D_H$, when written in terms of frequency, yields $\tau^{-1}_H\sim$\SI{2e12}{\per\second}, which is in the $10^{12}$-$10^{13}$~$s^{-1}$ range one would normally expect from phonons driving a thermally activated motion.
Note that this frequency is $\sim$5000 times smaller than what was found with \bnmr~\cite{2017-McFadden-CM-29-10187}, as well as with optical absorption~\cite{1964-Johnson-PR-136-A284}, which infer $D$ indirectly, whereas this is a direct measurement.  

  
In summary, we used the radioactive \mbox{$\alpha$-decay} of \eli\ to study \ce{Li} diffusion in a single crystal rutile \ce{TiO2} in the range of \SIrange{60}{370}{\kelvin}. 
The nanoscale Li diffusion rate was found to exhibit bi-Arrhenius behavior.
We report a high-$T$ activation energy of $E_{H}=$ \SI{0.3341 \pm  0.0021}{\electronvolt}, in agreement with measurements carried out with different techniques~\cite{1964-Johnson-PR-136-A284, 2010-Bach-EA-55-4952, 2017-McFadden-CM-29-10187}.
At low temperatures, a second Arrhenius component was revealed, with an activation energy of $E_L = \SI{0.0313 \pm 0.0015}{\electronvolt}$.
We suggest that this might be related to a small fraction of the \lip\ that does not bind to a Li-polaron complex but rather hops as a simple interstitial with an activation energy near theoretical calculations.
In addition, we found evidence that \ce{Li} traps at the $(001)$ surface, which could contribute to the reduced \ce{Li} uptake at room temperature.
We believe that this technique can shed new light on the \ce{Li} motion in Li-ion battery materials and across their interfaces. 

\begin{acknowledgments}
Special thanks to R. Abasalti, D. Vyas and M. McLay for all their excellent technical support.
This work was supported by: NSERC Discovery Grants to R.F.K. and W.A.M.;  J.S. was supported by Japan Society for the Promotion Science (JSPS) KAKENHI Grant No. JP18H01863; and \href{http://isosim.ubc.ca/}{IsoSiM} fellowships to A.C. and R.M.L.M.
TRIUMF receives federal funding via a contribution agreement with the National Research Council of Canada. 
\end{acknowledgments}

\bibliography{./references/refs,./references/alpha-tracer,./references/rutile,./references/misc,./references/geant4,./references/bnmr,./references/lithium8,./references/prefactors}

\begin{thebibliography}{38}%
\makeatletter
\providecommand \@ifxundefined [1]{%
 \@ifx{#1\undefined}
}%
\providecommand \@ifnum [1]{%
 \ifnum #1\expandafter \@firstoftwo
 \else \expandafter \@secondoftwo
 \fi
}%
\providecommand \@ifx [1]{%
 \ifx #1\expandafter \@firstoftwo
 \else \expandafter \@secondoftwo
 \fi
}%
\providecommand \natexlab [1]{#1}%
\providecommand \enquote  [1]{``#1''}%
\providecommand \bibnamefont  [1]{#1}%
\providecommand \bibfnamefont [1]{#1}%
\providecommand \citenamefont [1]{#1}%
\providecommand \href@noop [0]{\@secondoftwo}%
\providecommand \href [0]{\begingroup \@sanitize@url \@href}%
\providecommand \@href[1]{\@@startlink{#1}\@@href}%
\providecommand \@@href[1]{\endgroup#1\@@endlink}%
\providecommand \@sanitize@url [0]{\catcode `\\12\catcode `\$12\catcode
  `\&12\catcode `\#12\catcode `\^12\catcode `\_12\catcode `\%12\relax}%
\providecommand \@@startlink[1]{}%
\providecommand \@@endlink[0]{}%
\providecommand \url  [0]{\begingroup\@sanitize@url \@url }%
\providecommand \@url [1]{\endgroup\@href {#1}{\urlprefix }}%
\providecommand \urlprefix  [0]{URL }%
\providecommand \Eprint [0]{\href }%
\providecommand \doibase [0]{http://dx.doi.org/}%
\providecommand \selectlanguage [0]{\@gobble}%
\providecommand \bibinfo  [0]{\@secondoftwo}%
\providecommand \bibfield  [0]{\@secondoftwo}%
\providecommand \translation [1]{[#1]}%
\providecommand \BibitemOpen [0]{}%
\providecommand \bibitemStop [0]{}%
\providecommand \bibitemNoStop [0]{.\EOS\space}%
\providecommand \EOS [0]{\spacefactor3000\relax}%
\providecommand \BibitemShut  [1]{\csname bibitem#1\endcsname}%
\let\auto@bib@innerbib\@empty
\bibitem [{\citenamefont {Johnson}(1964)}]{1964-Johnson-PR-136-A284}%
  \BibitemOpen
  \bibfield  {author} {\bibinfo {author} {\bibfnamefont {O.~W.}\ \bibnamefont
  {Johnson}},\ }\href {\doibase 10.1103/PhysRev.136.A284} {\bibfield  {journal}
  {\bibinfo  {journal} {Phys. Rev.}\ }\textbf {\bibinfo {volume} {136}},\
  \bibinfo {pages} {A284} (\bibinfo {year} {1964})}\BibitemShut {NoStop}%
\bibitem [{\citenamefont {Johnson}\ and\ \citenamefont
  {Krouse}(1966)}]{1966-Johnson-JAP-37-668}%
  \BibitemOpen
  \bibfield  {author} {\bibinfo {author} {\bibfnamefont {O.~W.}\ \bibnamefont
  {Johnson}}\ and\ \bibinfo {author} {\bibfnamefont {H.~R.}\ \bibnamefont
  {Krouse}},\ }\href {\doibase 10.1063/1.1708234} {\bibfield  {journal}
  {\bibinfo  {journal} {J. Appl. Phys.}\ }\textbf {\bibinfo {volume} {37}},\
  \bibinfo {pages} {668} (\bibinfo {year} {1966})}\BibitemShut {NoStop}%
\bibitem [{\citenamefont {Van~Orman}\ and\ \citenamefont
  {Crispin}(2010)}]{2010-VanOrman-RMG-72-757}%
  \BibitemOpen
  \bibfield  {author} {\bibinfo {author} {\bibfnamefont {J.~A.}\ \bibnamefont
  {Van~Orman}}\ and\ \bibinfo {author} {\bibfnamefont {K.~L.}\ \bibnamefont
  {Crispin}},\ }\href {\doibase 10.2138/rmg.2010.72.17} {\bibfield  {journal}
  {\bibinfo  {journal} {Rev. Mineral. Geochem.}\ }\textbf {\bibinfo {volume}
  {72}},\ \bibinfo {pages} {757} (\bibinfo {year} {2010})}\BibitemShut
  {NoStop}%
\bibitem [{\citenamefont {Bachman}\ \emph {et~al.}(2016)\citenamefont
  {Bachman}, \citenamefont {Muy}, \citenamefont {Grimaud}, \citenamefont
  {Chang}, \citenamefont {Pour}, \citenamefont {Lux}, \citenamefont {Paschos},
  \citenamefont {Maglia}, \citenamefont {Lupart}, \citenamefont {Lamp},
  \citenamefont {Giordano},\ and\ \citenamefont
  {Shao-Horn}}]{2016-Bachman-CR-116-140}%
  \BibitemOpen
  \bibfield  {author} {\bibinfo {author} {\bibfnamefont {J.~C.}\ \bibnamefont
  {Bachman}}, \bibinfo {author} {\bibfnamefont {S.}~\bibnamefont {Muy}},
  \bibinfo {author} {\bibfnamefont {A.}~\bibnamefont {Grimaud}}, \bibinfo
  {author} {\bibfnamefont {H.-H.}\ \bibnamefont {Chang}}, \bibinfo {author}
  {\bibfnamefont {N.}~\bibnamefont {Pour}}, \bibinfo {author} {\bibfnamefont
  {S.~F.}\ \bibnamefont {Lux}}, \bibinfo {author} {\bibfnamefont
  {O.}~\bibnamefont {Paschos}}, \bibinfo {author} {\bibfnamefont
  {F.}~\bibnamefont {Maglia}}, \bibinfo {author} {\bibfnamefont
  {S.}~\bibnamefont {Lupart}}, \bibinfo {author} {\bibfnamefont
  {P.}~\bibnamefont {Lamp}}, \bibinfo {author} {\bibfnamefont {L.}~\bibnamefont
  {Giordano}}, \ and\ \bibinfo {author} {\bibfnamefont {Y.}~\bibnamefont
  {Shao-Horn}},\ }\href {\doibase 10.1021/acs.chemrev.5b00563} {\bibfield
  {journal} {\bibinfo  {journal} {Chem. Rev.}\ }\textbf {\bibinfo {volume}
  {116}},\ \bibinfo {pages} {140} (\bibinfo {year} {2016})}\BibitemShut
  {NoStop}%
\bibitem [{\citenamefont {Murphy}\ \emph {et~al.}(1983)\citenamefont {Murphy},
  \citenamefont {Cava}, \citenamefont {Zahurak},\ and\ \citenamefont
  {Santoro}}]{MURPHY1983413}%
  \BibitemOpen
  \bibfield  {author} {\bibinfo {author} {\bibfnamefont {D.~W.}\ \bibnamefont
  {Murphy}}, \bibinfo {author} {\bibfnamefont {R.~J.}\ \bibnamefont {Cava}},
  \bibinfo {author} {\bibfnamefont {S.~M.}\ \bibnamefont {Zahurak}}, \ and\
  \bibinfo {author} {\bibfnamefont {A.}~\bibnamefont {Santoro}},\ }\href
  {\doibase 10.1016/0167-2738(83)90268-0} {\bibfield  {journal} {\bibinfo
  {journal} {Solid State Ionics}\ }\textbf {\bibinfo {volume} {9--10}},\
  \bibinfo {pages} {413} (\bibinfo {year} {1983})}\BibitemShut {NoStop}%
\bibitem [{\citenamefont {Zachau-Christiansen}\ \emph
  {et~al.}(1988)\citenamefont {Zachau-Christiansen}, \citenamefont {West},
  \citenamefont {Jacobsen},\ and\ \citenamefont
  {Atlung}}]{ZACHAUCHRISTIANSEN19881176}%
  \BibitemOpen
  \bibfield  {author} {\bibinfo {author} {\bibfnamefont {B.}~\bibnamefont
  {Zachau-Christiansen}}, \bibinfo {author} {\bibfnamefont {K.}~\bibnamefont
  {West}}, \bibinfo {author} {\bibfnamefont {T.}~\bibnamefont {Jacobsen}}, \
  and\ \bibinfo {author} {\bibfnamefont {S.}~\bibnamefont {Atlung}},\ }\href
  {\doibase 10.1016/0167-2738(88)90352-9} {\bibfield  {journal} {\bibinfo
  {journal} {Solid State Ionics}\ }\textbf {\bibinfo {volume} {28--30}},\
  \bibinfo {pages} {1176} (\bibinfo {year} {1988})}\BibitemShut {NoStop}%
\bibitem [{\citenamefont {Hu}\ \emph {et~al.}(2006)\citenamefont {Hu},
  \citenamefont {Kienle}, \citenamefont {Guo},\ and\ \citenamefont
  {Maier}}]{2006-Hu-AM-18-1521}%
  \BibitemOpen
  \bibfield  {author} {\bibinfo {author} {\bibfnamefont {Y.-S.}\ \bibnamefont
  {Hu}}, \bibinfo {author} {\bibfnamefont {L.}~\bibnamefont {Kienle}}, \bibinfo
  {author} {\bibfnamefont {Y.-G.}\ \bibnamefont {Guo}}, \ and\ \bibinfo
  {author} {\bibfnamefont {J.}~\bibnamefont {Maier}},\ }\href {\doibase
  10.1002/adma.200502723} {\bibfield  {journal} {\bibinfo  {journal} {Adv.
  Mater.}\ }\textbf {\bibinfo {volume} {18}},\ \bibinfo {pages} {1421}
  (\bibinfo {year} {2006})}\BibitemShut {NoStop}%
\bibitem [{\citenamefont {Reddy}\ \emph {et~al.}(2013)\citenamefont {Reddy},
  \citenamefont {Subba~Rao},\ and\ \citenamefont
  {Chowdari}}]{2013-Reddy-CR-113-5364}%
  \BibitemOpen
  \bibfield  {author} {\bibinfo {author} {\bibfnamefont {M.~V.}\ \bibnamefont
  {Reddy}}, \bibinfo {author} {\bibfnamefont {G.~V.}\ \bibnamefont
  {Subba~Rao}}, \ and\ \bibinfo {author} {\bibfnamefont {B.~V.~R.}\
  \bibnamefont {Chowdari}},\ }\href {\doibase 10.1021/cr3001884} {\bibfield
  {journal} {\bibinfo  {journal} {Chem. Rev.}\ }\textbf {\bibinfo {volume}
  {113}},\ \bibinfo {pages} {5364} (\bibinfo {year} {2013})}\BibitemShut
  {NoStop}%
\bibitem [{\citenamefont {Kanamura}\ \emph {et~al.}(1987)\citenamefont
  {Kanamura}, \citenamefont {Yuasa},\ and\ \citenamefont
  {Takehara}}]{1987-Kanamura-JPS-20-127}%
  \BibitemOpen
  \bibfield  {author} {\bibinfo {author} {\bibfnamefont {K.}~\bibnamefont
  {Kanamura}}, \bibinfo {author} {\bibfnamefont {K.}~\bibnamefont {Yuasa}}, \
  and\ \bibinfo {author} {\bibfnamefont {Z.}~\bibnamefont {Takehara}},\ }\href
  {\doibase 10.1016/0378-7753(87)80101-5} {\bibfield  {journal} {\bibinfo
  {journal} {J. Power Sources}\ }\textbf {\bibinfo {volume} {20}},\ \bibinfo
  {pages} {127} (\bibinfo {year} {1987})}\BibitemShut {NoStop}%
\bibitem [{\citenamefont {Churikov}\ \emph {et~al.}(2004)\citenamefont
  {Churikov}, \citenamefont {Zobenkova},\ and\ \citenamefont
  {Pridatko}}]{2004-Churikov-RJE-40-63}%
  \BibitemOpen
  \bibfield  {author} {\bibinfo {author} {\bibfnamefont {A.~V.}\ \bibnamefont
  {Churikov}}, \bibinfo {author} {\bibfnamefont {V.~A.}\ \bibnamefont
  {Zobenkova}}, \ and\ \bibinfo {author} {\bibfnamefont {K.~I.}\ \bibnamefont
  {Pridatko}},\ }\href {\doibase 10.1023/B:RUEL.0000012076.90564.91} {\bibfield
   {journal} {\bibinfo  {journal} {Russ. J. Electrochem.}\ }\textbf {\bibinfo
  {volume} {40}},\ \bibinfo {pages} {63} (\bibinfo {year} {2004})}\BibitemShut
  {NoStop}%
\bibitem [{\citenamefont {Bach}\ \emph {et~al.}(2010)\citenamefont {Bach},
  \citenamefont {Pereira-Ramos},\ and\ \citenamefont
  {Willman}}]{2010-Bach-EA-55-4952}%
  \BibitemOpen
  \bibfield  {author} {\bibinfo {author} {\bibfnamefont {S.}~\bibnamefont
  {Bach}}, \bibinfo {author} {\bibfnamefont {J.~P.}\ \bibnamefont
  {Pereira-Ramos}}, \ and\ \bibinfo {author} {\bibfnamefont {P.}~\bibnamefont
  {Willman}},\ }\href {\doibase 10.1016/j.electacta.2010.03.101} {\bibfield
  {journal} {\bibinfo  {journal} {Electrochim. Acta.}\ }\textbf {\bibinfo
  {volume} {55}},\ \bibinfo {pages} {4952} (\bibinfo {year}
  {2010})}\BibitemShut {NoStop}%
\bibitem [{\citenamefont {Churikov}\ \emph {et~al.}(2014)\citenamefont
  {Churikov}, \citenamefont {Ivanishchev}, \citenamefont {Ushakov},\ and\
  \citenamefont {Romanova}}]{2014-Churikov-JSSE-18-1425}%
  \BibitemOpen
  \bibfield  {author} {\bibinfo {author} {\bibfnamefont {A.~V.}\ \bibnamefont
  {Churikov}}, \bibinfo {author} {\bibfnamefont {A.~V.}\ \bibnamefont
  {Ivanishchev}}, \bibinfo {author} {\bibfnamefont {A.~V.}\ \bibnamefont
  {Ushakov}}, \ and\ \bibinfo {author} {\bibfnamefont {V.~O.}\ \bibnamefont
  {Romanova}},\ }\href {\doibase 10.1007/s10008-013-2358-y} {\bibfield
  {journal} {\bibinfo  {journal} {J. Solid State Electrochem.}\ }\textbf
  {\bibinfo {volume} {18}},\ \bibinfo {pages} {1425} (\bibinfo {year}
  {2014})}\BibitemShut {NoStop}%
\bibitem [{\citenamefont {McFadden}\ \emph {et~al.}(2017)\citenamefont
  {McFadden}, \citenamefont {Buck}, \citenamefont {Chatzichristos},
  \citenamefont {Chen}, \citenamefont {Chow}, \citenamefont {Cortie},
  \citenamefont {Dehn}, \citenamefont {Karner}, \citenamefont {Koumoulis},
  \citenamefont {Levy}, \citenamefont {Li}, \citenamefont {McKenzie},
  \citenamefont {Merkle}, \citenamefont {Morris}, \citenamefont {Pearson},
  \citenamefont {Salman}, \citenamefont {Samuelis}, \citenamefont {Stachura},
  \citenamefont {Xiao}, \citenamefont {Maier}, \citenamefont {Kiefl},\ and\
  \citenamefont {MacFarlane}}]{2017-McFadden-CM-29-10187}%
  \BibitemOpen
  \bibfield  {author} {\bibinfo {author} {\bibfnamefont {R.~M.~L.}\
  \bibnamefont {McFadden}}, \bibinfo {author} {\bibfnamefont {T.~J.}\
  \bibnamefont {Buck}}, \bibinfo {author} {\bibfnamefont {A.}~\bibnamefont
  {Chatzichristos}}, \bibinfo {author} {\bibfnamefont {C.-C.}\ \bibnamefont
  {Chen}}, \bibinfo {author} {\bibfnamefont {K.~H.}\ \bibnamefont {Chow}},
  \bibinfo {author} {\bibfnamefont {D.~L.}\ \bibnamefont {Cortie}}, \bibinfo
  {author} {\bibfnamefont {M.~H.}\ \bibnamefont {Dehn}}, \bibinfo {author}
  {\bibfnamefont {V.~L.}\ \bibnamefont {Karner}}, \bibinfo {author}
  {\bibfnamefont {D.}~\bibnamefont {Koumoulis}}, \bibinfo {author}
  {\bibfnamefont {C.~D.~P.}\ \bibnamefont {Levy}}, \bibinfo {author}
  {\bibfnamefont {C.}~\bibnamefont {Li}}, \bibinfo {author} {\bibfnamefont
  {I.}~\bibnamefont {McKenzie}}, \bibinfo {author} {\bibfnamefont
  {R.}~\bibnamefont {Merkle}}, \bibinfo {author} {\bibfnamefont {G.~D.}\
  \bibnamefont {Morris}}, \bibinfo {author} {\bibfnamefont {M.~R.}\
  \bibnamefont {Pearson}}, \bibinfo {author} {\bibfnamefont {Z.}~\bibnamefont
  {Salman}}, \bibinfo {author} {\bibfnamefont {D.}~\bibnamefont {Samuelis}},
  \bibinfo {author} {\bibfnamefont {M.}~\bibnamefont {Stachura}}, \bibinfo
  {author} {\bibfnamefont {J.}~\bibnamefont {Xiao}}, \bibinfo {author}
  {\bibfnamefont {J.}~\bibnamefont {Maier}}, \bibinfo {author} {\bibfnamefont
  {R.~F.}\ \bibnamefont {Kiefl}}, \ and\ \bibinfo {author} {\bibfnamefont
  {W.~A.}\ \bibnamefont {MacFarlane}},\ }\href {\doibase
  10.1021/acs.chemmater.7b04093} {\bibfield  {journal} {\bibinfo  {journal}
  {Chem. Mater.}\ }\textbf {\bibinfo {volume} {29}},\ \bibinfo {pages} {10187}
  (\bibinfo {year} {2017})}\BibitemShut {NoStop}%
\bibitem [{\citenamefont {Koudriachova}\ \emph {et~al.}(2002)\citenamefont
  {Koudriachova}, \citenamefont {Harrison},\ and\ \citenamefont
  {de~Leeuw}}]{2002-Koudriachova-PRB-65-235423}%
  \BibitemOpen
  \bibfield  {author} {\bibinfo {author} {\bibfnamefont {M.~V.}\ \bibnamefont
  {Koudriachova}}, \bibinfo {author} {\bibfnamefont {N.~M.}\ \bibnamefont
  {Harrison}}, \ and\ \bibinfo {author} {\bibfnamefont {S.~W.}\ \bibnamefont
  {de~Leeuw}},\ }\href {\doibase 10.1103/PhysRevB.65.235423} {\bibfield
  {journal} {\bibinfo  {journal} {Phys. Rev. B}\ }\textbf {\bibinfo {volume}
  {65}},\ \bibinfo {pages} {235423} (\bibinfo {year} {2002})}\BibitemShut
  {NoStop}%
\bibitem [{\citenamefont {Gligor}\ and\ \citenamefont
  {de~Leeuw}(2006)}]{2006-Gligor-SSI-177-2741}%
  \BibitemOpen
  \bibfield  {author} {\bibinfo {author} {\bibfnamefont {F.}~\bibnamefont
  {Gligor}}\ and\ \bibinfo {author} {\bibfnamefont {S.~W.}\ \bibnamefont
  {de~Leeuw}},\ }\href {\doibase 10.1016/j.ssi.2006.03.017} {\bibfield
  {journal} {\bibinfo  {journal} {Solid State Ionics}\ }\textbf {\bibinfo
  {volume} {177}},\ \bibinfo {pages} {2741} (\bibinfo {year}
  {2006})}\BibitemShut {NoStop}%
\bibitem [{\citenamefont {Kerisit}\ \emph
  {et~al.}(2009{\natexlab{a}})\citenamefont {Kerisit}, \citenamefont {Rosso},
  \citenamefont {Yang},\ and\ \citenamefont
  {Liu}}]{2009-Kerisit-JPCC-113-20998}%
  \BibitemOpen
  \bibfield  {author} {\bibinfo {author} {\bibfnamefont {S.}~\bibnamefont
  {Kerisit}}, \bibinfo {author} {\bibfnamefont {K.~M.}\ \bibnamefont {Rosso}},
  \bibinfo {author} {\bibfnamefont {Z.}~\bibnamefont {Yang}}, \ and\ \bibinfo
  {author} {\bibfnamefont {J.}~\bibnamefont {Liu}},\ }\href {\doibase
  10.1021/jp9064517} {\bibfield  {journal} {\bibinfo  {journal} {J. Phys. Chem.
  C}\ }\textbf {\bibinfo {volume} {113}},\ \bibinfo {pages} {20998} (\bibinfo
  {year} {2009}{\natexlab{a}})}\BibitemShut {NoStop}%
\bibitem [{\citenamefont {Yildirim}\ \emph {et~al.}(2012)\citenamefont
  {Yildirim}, \citenamefont {Greeley},\ and\ \citenamefont
  {Sankaranarayanan}}]{2012-Yildirim-PCCP-14-4565}%
  \BibitemOpen
  \bibfield  {author} {\bibinfo {author} {\bibfnamefont {H.}~\bibnamefont
  {Yildirim}}, \bibinfo {author} {\bibfnamefont {J.~P.}\ \bibnamefont
  {Greeley}}, \ and\ \bibinfo {author} {\bibfnamefont {S.~K. R.~S.}\
  \bibnamefont {Sankaranarayanan}},\ }\href {\doibase 10.1039/C2CP22731B}
  {\bibfield  {journal} {\bibinfo  {journal} {Phys. Chem. Chem. Phys.}\
  }\textbf {\bibinfo {volume} {14}},\ \bibinfo {pages} {4565} (\bibinfo {year}
  {2012})}\BibitemShut {NoStop}%
\bibitem [{\citenamefont {Kerisit}\ \emph {et~al.}(2014)\citenamefont
  {Kerisit}, \citenamefont {Chaka}, \citenamefont {Droubay},\ and\
  \citenamefont {Ilton}}]{2014-Kerisit-JPCC-118-24231}%
  \BibitemOpen
  \bibfield  {author} {\bibinfo {author} {\bibfnamefont {S.}~\bibnamefont
  {Kerisit}}, \bibinfo {author} {\bibfnamefont {A.~M.}\ \bibnamefont {Chaka}},
  \bibinfo {author} {\bibfnamefont {T.~C.}\ \bibnamefont {Droubay}}, \ and\
  \bibinfo {author} {\bibfnamefont {E.~S.}\ \bibnamefont {Ilton}},\ }\href
  {\doibase 10.1021/jp506025k} {\bibfield  {journal} {\bibinfo  {journal} {J.
  Phys. Chem. C}\ }\textbf {\bibinfo {volume} {118}},\ \bibinfo {pages} {24231}
  (\bibinfo {year} {2014})}\BibitemShut {NoStop}%
\bibitem [{\citenamefont {Jung}\ \emph {et~al.}(2014)\citenamefont {Jung},
  \citenamefont {Cho},\ and\ \citenamefont {Zhou}}]{2014-Jung-AIPA-4-017104}%
  \BibitemOpen
  \bibfield  {author} {\bibinfo {author} {\bibfnamefont {J.}~\bibnamefont
  {Jung}}, \bibinfo {author} {\bibfnamefont {M.}~\bibnamefont {Cho}}, \ and\
  \bibinfo {author} {\bibfnamefont {M.}~\bibnamefont {Zhou}},\ }\href {\doibase
  10.1063/1.4861583} {\bibfield  {journal} {\bibinfo  {journal} {AIP Adv.}\
  }\textbf {\bibinfo {volume} {4}},\ \bibinfo {pages} {017104} (\bibinfo {year}
  {2014})}\BibitemShut {NoStop}%
\bibitem [{\citenamefont {Arrouvel}\ \emph {et~al.}(2015)\citenamefont
  {Arrouvel}, \citenamefont {Peixoto}, \citenamefont {Valerio},\ and\
  \citenamefont {Parker}}]{2015-Arrouvel-CTC-1072-43}%
  \BibitemOpen
  \bibfield  {author} {\bibinfo {author} {\bibfnamefont {C.}~\bibnamefont
  {Arrouvel}}, \bibinfo {author} {\bibfnamefont {T.~C.}\ \bibnamefont
  {Peixoto}}, \bibinfo {author} {\bibfnamefont {M.~E.~G.}\ \bibnamefont
  {Valerio}}, \ and\ \bibinfo {author} {\bibfnamefont {S.~C.}\ \bibnamefont
  {Parker}},\ }\href {\doibase 10.1016/j.comptc.2015.09.002} {\bibfield
  {journal} {\bibinfo  {journal} {Comput. Theor. Chem.}\ }\textbf {\bibinfo
  {volume} {1072}},\ \bibinfo {pages} {43} (\bibinfo {year}
  {2015})}\BibitemShut {NoStop}%
\bibitem [{\citenamefont {Ishiyama}\ \emph {et~al.}(2016)\citenamefont
  {Ishiyama}, \citenamefont {Jeong}, \citenamefont {Watanabe}, \citenamefont
  {Hirayama}, \citenamefont {Imai}, \citenamefont {Jung}, \citenamefont
  {Miyatake}, \citenamefont {Oyaizu}, \citenamefont {Osa}, \citenamefont
  {Otokawa}, \citenamefont {Matsuda}, \citenamefont {Nishio}, \citenamefont
  {Makii}, \citenamefont {Sato}, \citenamefont {Kuwata}, \citenamefont
  {Kawamura}, \citenamefont {Ueno}, \citenamefont {Kim}, \citenamefont
  {Kimura},\ and\ \citenamefont {Mukai}}]{2016-Ishiyama-NIMB-376-379}%
  \BibitemOpen
  \bibfield  {author} {\bibinfo {author} {\bibfnamefont {H.}~\bibnamefont
  {Ishiyama}}, \bibinfo {author} {\bibfnamefont {S.~C.}\ \bibnamefont {Jeong}},
  \bibinfo {author} {\bibfnamefont {Y.~X.}\ \bibnamefont {Watanabe}}, \bibinfo
  {author} {\bibfnamefont {Y.}~\bibnamefont {Hirayama}}, \bibinfo {author}
  {\bibfnamefont {N.}~\bibnamefont {Imai}}, \bibinfo {author} {\bibfnamefont
  {H.~S.}\ \bibnamefont {Jung}}, \bibinfo {author} {\bibfnamefont
  {H.}~\bibnamefont {Miyatake}}, \bibinfo {author} {\bibfnamefont
  {M.}~\bibnamefont {Oyaizu}}, \bibinfo {author} {\bibfnamefont
  {A.}~\bibnamefont {Osa}}, \bibinfo {author} {\bibfnamefont {Y.}~\bibnamefont
  {Otokawa}}, \bibinfo {author} {\bibfnamefont {M.}~\bibnamefont {Matsuda}},
  \bibinfo {author} {\bibfnamefont {K.}~\bibnamefont {Nishio}}, \bibinfo
  {author} {\bibfnamefont {H.}~\bibnamefont {Makii}}, \bibinfo {author}
  {\bibfnamefont {T.~K.}\ \bibnamefont {Sato}}, \bibinfo {author}
  {\bibfnamefont {N.}~\bibnamefont {Kuwata}}, \bibinfo {author} {\bibfnamefont
  {J.}~\bibnamefont {Kawamura}}, \bibinfo {author} {\bibfnamefont
  {H.}~\bibnamefont {Ueno}}, \bibinfo {author} {\bibfnamefont {Y.~H.}\
  \bibnamefont {Kim}}, \bibinfo {author} {\bibfnamefont {S.}~\bibnamefont
  {Kimura}}, \ and\ \bibinfo {author} {\bibfnamefont {M.}~\bibnamefont
  {Mukai}},\ }\href {\doibase 10.1016/j.nimb.2015.12.036} {\bibfield  {journal}
  {\bibinfo  {journal} {Nucl. Instrum. Methods Phys. Res., Sect. B}\ }\textbf
  {\bibinfo {volume} {376}},\ \bibinfo {pages} {379} (\bibinfo {year}
  {2016})}\BibitemShut {NoStop}%
\bibitem [{\citenamefont {Morris}(2014)}]{2014-Morris-HI-225-173}%
  \BibitemOpen
  \bibfield  {author} {\bibinfo {author} {\bibfnamefont {G.~D.}\ \bibnamefont
  {Morris}},\ }\href {\doibase 10.1007/s10751-013-0894-6} {\bibfield  {journal}
  {\bibinfo  {journal} {Hyperfine Interact.}\ }\textbf {\bibinfo {volume}
  {225}},\ \bibinfo {pages} {173} (\bibinfo {year} {2014})}\BibitemShut
  {NoStop}%
\bibitem [{\citenamefont {Salman}\ \emph {et~al.}(2004)\citenamefont {Salman},
  \citenamefont {Reynard}, \citenamefont {MacFarlane}, \citenamefont {Chow},
  \citenamefont {Chakhalian}, \citenamefont {Kreitzman}, \citenamefont
  {Daviel}, \citenamefont {Levy}, \citenamefont {Poutissou},\ and\
  \citenamefont {Kiefl}}]{2004-Salman-PRB-70-104404}%
  \BibitemOpen
  \bibfield  {author} {\bibinfo {author} {\bibfnamefont {Z.}~\bibnamefont
  {Salman}}, \bibinfo {author} {\bibfnamefont {E.~P.}\ \bibnamefont {Reynard}},
  \bibinfo {author} {\bibfnamefont {W.~A.}\ \bibnamefont {MacFarlane}},
  \bibinfo {author} {\bibfnamefont {K.~H.}\ \bibnamefont {Chow}}, \bibinfo
  {author} {\bibfnamefont {J.}~\bibnamefont {Chakhalian}}, \bibinfo {author}
  {\bibfnamefont {S.~R.}\ \bibnamefont {Kreitzman}}, \bibinfo {author}
  {\bibfnamefont {S.}~\bibnamefont {Daviel}}, \bibinfo {author} {\bibfnamefont
  {C.~D.~P.}\ \bibnamefont {Levy}}, \bibinfo {author} {\bibfnamefont
  {R.}~\bibnamefont {Poutissou}}, \ and\ \bibinfo {author} {\bibfnamefont
  {R.~F.}\ \bibnamefont {Kiefl}},\ }\href {\doibase 10.1103/PhysRevB.70.104404}
  {\bibfield  {journal} {\bibinfo  {journal} {Phys. Rev. B}\ }\textbf {\bibinfo
  {volume} {70}},\ \bibinfo {pages} {104404} (\bibinfo {year}
  {2004})}\BibitemShut {NoStop}%
\bibitem [{\citenamefont {Asada}\ \emph {et~al.}(1959)\citenamefont {Asada},
  \citenamefont {Masuda}, \citenamefont {Okumura},\ and\ \citenamefont
  {Okuma}}]{1959-Asada-JPSJ-14-12}%
  \BibitemOpen
  \bibfield  {author} {\bibinfo {author} {\bibfnamefont {T.}~\bibnamefont
  {Asada}}, \bibinfo {author} {\bibfnamefont {M.}~\bibnamefont {Masuda}},
  \bibinfo {author} {\bibfnamefont {M.}~\bibnamefont {Okumura}}, \ and\
  \bibinfo {author} {\bibfnamefont {J.}~\bibnamefont {Okuma}},\ }\href
  {\doibase 10.1143/JPSJ.14.1766} {\bibfield  {journal} {\bibinfo  {journal}
  {Journal of the Physical Society of Japan}\ }\textbf {\bibinfo {volume}
  {14}},\ \bibinfo {pages} {1766} (\bibinfo {year} {1959})}\BibitemShut
  {NoStop}%
\bibitem [{\citenamefont {Ishiyama}\ \emph {et~al.}(2013)\citenamefont
  {Ishiyama}, \citenamefont {Jeong}, \citenamefont {Watanabe}, \citenamefont
  {Hirayama}, \citenamefont {Imai}, \citenamefont {Miyatake}, \citenamefont
  {Oyaizu}, \citenamefont {Katayama}, \citenamefont {Sataka}, \citenamefont
  {Osa}, \citenamefont {Otokawa}, \citenamefont {Matsuda},\ and\ \citenamefont
  {Makii}}]{2013-Ishiyama-JJAP-52-010205}%
  \BibitemOpen
  \bibfield  {author} {\bibinfo {author} {\bibfnamefont {H.}~\bibnamefont
  {Ishiyama}}, \bibinfo {author} {\bibfnamefont {S.-C.}\ \bibnamefont {Jeong}},
  \bibinfo {author} {\bibfnamefont {Y.}~\bibnamefont {Watanabe}}, \bibinfo
  {author} {\bibfnamefont {Y.}~\bibnamefont {Hirayama}}, \bibinfo {author}
  {\bibfnamefont {N.}~\bibnamefont {Imai}}, \bibinfo {author} {\bibfnamefont
  {H.}~\bibnamefont {Miyatake}}, \bibinfo {author} {\bibfnamefont
  {M.}~\bibnamefont {Oyaizu}}, \bibinfo {author} {\bibfnamefont
  {I.}~\bibnamefont {Katayama}}, \bibinfo {author} {\bibfnamefont
  {M.}~\bibnamefont {Sataka}}, \bibinfo {author} {\bibfnamefont
  {A.}~\bibnamefont {Osa}}, \bibinfo {author} {\bibfnamefont {Y.}~\bibnamefont
  {Otokawa}}, \bibinfo {author} {\bibfnamefont {M.}~\bibnamefont {Matsuda}}, \
  and\ \bibinfo {author} {\bibfnamefont {H.}~\bibnamefont {Makii}},\ }\href
  {\doibase 10.7567/JJAP.52.010205} {\bibfield  {journal} {\bibinfo  {journal}
  {Jpn. J. Appl. Phys.}\ }\textbf {\bibinfo {volume} {52}},\ \bibinfo {pages}
  {010205} (\bibinfo {year} {2013})}\BibitemShut {NoStop}%
\bibitem [{\citenamefont {Ishiyama}\ \emph {et~al.}(2015)\citenamefont
  {Ishiyama}, \citenamefont {Jeong}, \citenamefont {Watanabe}, \citenamefont
  {Hirayama}, \citenamefont {Imai}, \citenamefont {Miyatake}, \citenamefont
  {Oyaizu}, \citenamefont {Osa}, \citenamefont {Otokawa}, \citenamefont
  {Matsuda}, \citenamefont {Nishio}, \citenamefont {Makii}, \citenamefont
  {Sato}, \citenamefont {Kuwata}, \citenamefont {Kawamura}, \citenamefont
  {Nakao}, \citenamefont {Ueno}, \citenamefont {Kim}, \citenamefont {Kimura},\
  and\ \citenamefont {Mukai}}]{2015-Ishiyama-NIMB-354-297}%
  \BibitemOpen
  \bibfield  {author} {\bibinfo {author} {\bibfnamefont {H.}~\bibnamefont
  {Ishiyama}}, \bibinfo {author} {\bibfnamefont {S.~C.}\ \bibnamefont {Jeong}},
  \bibinfo {author} {\bibfnamefont {Y.~X.}\ \bibnamefont {Watanabe}}, \bibinfo
  {author} {\bibfnamefont {Y.}~\bibnamefont {Hirayama}}, \bibinfo {author}
  {\bibfnamefont {N.}~\bibnamefont {Imai}}, \bibinfo {author} {\bibfnamefont
  {H.}~\bibnamefont {Miyatake}}, \bibinfo {author} {\bibfnamefont
  {M.}~\bibnamefont {Oyaizu}}, \bibinfo {author} {\bibfnamefont
  {A.}~\bibnamefont {Osa}}, \bibinfo {author} {\bibfnamefont {Y.}~\bibnamefont
  {Otokawa}}, \bibinfo {author} {\bibfnamefont {M.}~\bibnamefont {Matsuda}},
  \bibinfo {author} {\bibfnamefont {K.}~\bibnamefont {Nishio}}, \bibinfo
  {author} {\bibfnamefont {H.}~\bibnamefont {Makii}}, \bibinfo {author}
  {\bibfnamefont {T.~K.}\ \bibnamefont {Sato}}, \bibinfo {author}
  {\bibfnamefont {N.}~\bibnamefont {Kuwata}}, \bibinfo {author} {\bibfnamefont
  {J.}~\bibnamefont {Kawamura}}, \bibinfo {author} {\bibfnamefont
  {A.}~\bibnamefont {Nakao}}, \bibinfo {author} {\bibfnamefont
  {H.}~\bibnamefont {Ueno}}, \bibinfo {author} {\bibfnamefont {Y.~H.}\
  \bibnamefont {Kim}}, \bibinfo {author} {\bibfnamefont {S.}~\bibnamefont
  {Kimura}}, \ and\ \bibinfo {author} {\bibfnamefont {M.}~\bibnamefont
  {Mukai}},\ }\href {\doibase 10.1016/j.nimb.2014.10.031} {\bibfield  {journal}
  {\bibinfo  {journal} {Nucl. Instrum. Methods Phys. Res., Sect. B}\ }\textbf
  {\bibinfo {volume} {354}},\ \bibinfo {pages} {297} (\bibinfo {year}
  {2015})}\BibitemShut {NoStop}%
\bibitem [{\citenamefont {{Ziegler}}\ \emph {et~al.}(2010)\citenamefont
  {{Ziegler}}, \citenamefont {{Ziegler}},\ and\ \citenamefont
  {{Biersack}}}]{SRIM}%
  \BibitemOpen
  \bibfield  {author} {\bibinfo {author} {\bibfnamefont {J.~F.}\ \bibnamefont
  {{Ziegler}}}, \bibinfo {author} {\bibfnamefont {M.~D.}\ \bibnamefont
  {{Ziegler}}}, \ and\ \bibinfo {author} {\bibfnamefont {J.~P.}\ \bibnamefont
  {{Biersack}}},\ }\href {\doibase 10.1016/j.nimb.2010.02.091} {\bibfield
  {journal} {\bibinfo  {journal} {Nucl. Instrum. Methods Phys. Res., Sect. B}\
  }\textbf {\bibinfo {volume} {268}},\ \bibinfo {pages} {1818} (\bibinfo {year}
  {2010})}\BibitemShut {NoStop}%
\bibitem [{\citenamefont {{J. Allison}}\ \emph {et~al.}(2016)\citenamefont {{J.
  Allison}} \emph {et~al.}}]{Geant4-dev-2016}%
  \BibitemOpen
  \bibfield  {author} {\bibinfo {author} {\bibnamefont {{J. Allison}}} \emph
  {et~al.},\ }\href {\doibase 10.1016/j.nima.2016.06.125} {\bibfield  {journal}
  {\bibinfo  {journal} {Nucl. Instrum. Methods Phys. Res., Sect. A}\ }\textbf
  {\bibinfo {volume} {835}},\ \bibinfo {pages} {186} (\bibinfo {year}
  {2016})}\BibitemShut {NoStop}%
\bibitem [{\citenamefont {Jeong}\ \emph {et~al.}(2005)\citenamefont {Jeong},
  \citenamefont {Katayama}, \citenamefont {Kawakami}, \citenamefont {Watanabe},
  \citenamefont {Ishiyama}, \citenamefont {Miyatake}, \citenamefont {Sataka},
  \citenamefont {Okayasu}, \citenamefont {Sugai}, \citenamefont {Ichikawa},
  \citenamefont {Nishio}, \citenamefont {Nakanoya}, \citenamefont {Ishikawa},
  \citenamefont {Chimi}, \citenamefont {Hashimoto}, \citenamefont {Yahagi},
  \citenamefont {Takada}, \citenamefont {Kim}, \citenamefont {Watanabe},
  \citenamefont {Iwase}, \citenamefont {Hashimoto},\ and\ \citenamefont
  {Ishikawa}}]{2005-Jeong-NIMB-230-596}%
  \BibitemOpen
  \bibfield  {author} {\bibinfo {author} {\bibfnamefont {S.-C.}\ \bibnamefont
  {Jeong}}, \bibinfo {author} {\bibfnamefont {I.}~\bibnamefont {Katayama}},
  \bibinfo {author} {\bibfnamefont {H.}~\bibnamefont {Kawakami}}, \bibinfo
  {author} {\bibfnamefont {Y.}~\bibnamefont {Watanabe}}, \bibinfo {author}
  {\bibfnamefont {H.}~\bibnamefont {Ishiyama}}, \bibinfo {author}
  {\bibfnamefont {H.}~\bibnamefont {Miyatake}}, \bibinfo {author}
  {\bibfnamefont {M.}~\bibnamefont {Sataka}}, \bibinfo {author} {\bibfnamefont
  {S.}~\bibnamefont {Okayasu}}, \bibinfo {author} {\bibfnamefont
  {H.}~\bibnamefont {Sugai}}, \bibinfo {author} {\bibfnamefont
  {S.}~\bibnamefont {Ichikawa}}, \bibinfo {author} {\bibfnamefont
  {K.}~\bibnamefont {Nishio}}, \bibinfo {author} {\bibfnamefont
  {T.}~\bibnamefont {Nakanoya}}, \bibinfo {author} {\bibfnamefont
  {N.}~\bibnamefont {Ishikawa}}, \bibinfo {author} {\bibfnamefont
  {Y.}~\bibnamefont {Chimi}}, \bibinfo {author} {\bibfnamefont
  {T.}~\bibnamefont {Hashimoto}}, \bibinfo {author} {\bibfnamefont
  {M.}~\bibnamefont {Yahagi}}, \bibinfo {author} {\bibfnamefont
  {K.}~\bibnamefont {Takada}}, \bibinfo {author} {\bibfnamefont {B.~C.}\
  \bibnamefont {Kim}}, \bibinfo {author} {\bibfnamefont {M.}~\bibnamefont
  {Watanabe}}, \bibinfo {author} {\bibfnamefont {A.}~\bibnamefont {Iwase}},
  \bibinfo {author} {\bibfnamefont {T.}~\bibnamefont {Hashimoto}}, \ and\
  \bibinfo {author} {\bibfnamefont {T.}~\bibnamefont {Ishikawa}},\ }\href
  {\doibase 10.1016/j.nimb.2004.12.107} {\bibfield  {journal} {\bibinfo
  {journal} {Nucl. Instrum. Methods Phys. Res., Sect. B}\ }\textbf {\bibinfo
  {volume} {230}},\ \bibinfo {pages} {596} (\bibinfo {year}
  {2005})}\BibitemShut {NoStop}%
\bibitem [{\citenamefont {James}\ and\ \citenamefont
  {Roos}(1975)}]{1975-James-CPC-10-343}%
  \BibitemOpen
  \bibfield  {author} {\bibinfo {author} {\bibfnamefont {F.}~\bibnamefont
  {James}}\ and\ \bibinfo {author} {\bibfnamefont {M.}~\bibnamefont {Roos}},\
  }\href {\doibase 10.1016/0010-4655(75)90039-9} {\bibfield  {journal}
  {\bibinfo  {journal} {Comput. Phys. Commun.}\ }\textbf {\bibinfo {volume}
  {10}},\ \bibinfo {pages} {343} (\bibinfo {year} {1975})}\BibitemShut
  {NoStop}%
\bibitem [{\citenamefont {Brun}\ and\ \citenamefont
  {Rademakers}(1997)}]{1997-Brun-NIMA-389-81}%
  \BibitemOpen
  \bibfield  {author} {\bibinfo {author} {\bibfnamefont {R.}~\bibnamefont
  {Brun}}\ and\ \bibinfo {author} {\bibfnamefont {F.}~\bibnamefont
  {Rademakers}},\ }\href {\doibase 10.1016/S0168-9002(97)00048-X} {\bibfield
  {journal} {\bibinfo  {journal} {Nucl. Instrum. Methods Phys. Res., Sect. A}\
  }\textbf {\bibinfo {volume} {389}},\ \bibinfo {pages} {81} (\bibinfo {year}
  {1997})}\BibitemShut {NoStop}%
\bibitem [{\citenamefont {Okuyama}\ \emph {et~al.}(1998)\citenamefont
  {Okuyama}, \citenamefont {Siga}, \citenamefont {Takagi}, \citenamefont
  {Nishijima},\ and\ \citenamefont {Aruga}}]{1998-Okuyama-SS-401-3}%
  \BibitemOpen
  \bibfield  {author} {\bibinfo {author} {\bibfnamefont {H.}~\bibnamefont
  {Okuyama}}, \bibinfo {author} {\bibfnamefont {W.}~\bibnamefont {Siga}},
  \bibinfo {author} {\bibfnamefont {N.}~\bibnamefont {Takagi}}, \bibinfo
  {author} {\bibfnamefont {M.}~\bibnamefont {Nishijima}}, \ and\ \bibinfo
  {author} {\bibfnamefont {T.}~\bibnamefont {Aruga}},\ }\href {\doibase
  https://doi.org/10.1016/S0039-6028(98)00020-X} {\bibfield  {journal}
  {\bibinfo  {journal} {Surface Science}\ }\textbf {\bibinfo {volume} {401}},\
  \bibinfo {pages} {344 } (\bibinfo {year} {1998})}\BibitemShut {NoStop}%
\bibitem [{\citenamefont {Diebold}(2003)}]{2003-Diebold-SSR-48-53}%
  \BibitemOpen
  \bibfield  {author} {\bibinfo {author} {\bibfnamefont {U.}~\bibnamefont
  {Diebold}},\ }\href {\doibase 10.1016/S0167-5729(02)00100-0} {\bibfield
  {journal} {\bibinfo  {journal} {Surf. Sci. Rep.}\ }\textbf {\bibinfo {volume}
  {48}},\ \bibinfo {pages} {53} (\bibinfo {year} {2003})}\BibitemShut {NoStop}%
\bibitem [{\citenamefont {Beals}\ \emph {et~al.}(2003)\citenamefont {Beals},
  \citenamefont {Kiefl}, \citenamefont {MacFarlane}, \citenamefont {Nichol},
  \citenamefont {Morris}, \citenamefont {Levy}, \citenamefont {Kreitzman},
  \citenamefont {Poutissou}, \citenamefont {Daviel}, \citenamefont {Baartman},\
  and\ \citenamefont {Chow}}]{2003-Beals-PB-326-1}%
  \BibitemOpen
  \bibfield  {author} {\bibinfo {author} {\bibfnamefont {T.}~\bibnamefont
  {Beals}}, \bibinfo {author} {\bibfnamefont {R.}~\bibnamefont {Kiefl}},
  \bibinfo {author} {\bibfnamefont {W.}~\bibnamefont {MacFarlane}}, \bibinfo
  {author} {\bibfnamefont {K.}~\bibnamefont {Nichol}}, \bibinfo {author}
  {\bibfnamefont {G.}~\bibnamefont {Morris}}, \bibinfo {author} {\bibfnamefont
  {C.}~\bibnamefont {Levy}}, \bibinfo {author} {\bibfnamefont {S.}~\bibnamefont
  {Kreitzman}}, \bibinfo {author} {\bibfnamefont {R.}~\bibnamefont
  {Poutissou}}, \bibinfo {author} {\bibfnamefont {S.}~\bibnamefont {Daviel}},
  \bibinfo {author} {\bibfnamefont {R.}~\bibnamefont {Baartman}}, \ and\
  \bibinfo {author} {\bibfnamefont {K.}~\bibnamefont {Chow}},\ }\href {\doibase
  10.1016/S0921-4526(02)01602-2} {\bibfield  {journal} {\bibinfo  {journal}
  {Physica B: Condensed Matter}\ }\textbf {\bibinfo {volume} {326}},\ \bibinfo
  {pages} {205 } (\bibinfo {year} {2003})}\BibitemShut {NoStop}%
\bibitem [{\citenamefont {Hooton}\ and\ \citenamefont
  {Jacobs}(1988)}]{1988-Hooton-CJC-66-4}%
  \BibitemOpen
  \bibfield  {author} {\bibinfo {author} {\bibfnamefont {I.~E.}\ \bibnamefont
  {Hooton}}\ and\ \bibinfo {author} {\bibfnamefont {P.~W.~M.}\ \bibnamefont
  {Jacobs}},\ }\href {\doibase 10.1139/v88-144} {\bibfield  {journal} {\bibinfo
   {journal} {Canadian Journal of Chemistry}\ }\textbf {\bibinfo {volume}
  {66}},\ \bibinfo {pages} {830} (\bibinfo {year} {1988})}\BibitemShut
  {NoStop}%
\bibitem [{\citenamefont {Kerisit}\ \emph
  {et~al.}(2009{\natexlab{b}})\citenamefont {Kerisit}, \citenamefont {Rosso},
  \citenamefont {Yang},\ and\ \citenamefont {Liu}}]{2009-Kerisit-JPCC-113-49}%
  \BibitemOpen
  \bibfield  {author} {\bibinfo {author} {\bibfnamefont {S.}~\bibnamefont
  {Kerisit}}, \bibinfo {author} {\bibfnamefont {K.~M.}\ \bibnamefont {Rosso}},
  \bibinfo {author} {\bibfnamefont {Z.}~\bibnamefont {Yang}}, \ and\ \bibinfo
  {author} {\bibfnamefont {J.}~\bibnamefont {Liu}},\ }\href {\doibase
  10.1021/jp9064517} {\bibfield  {journal} {\bibinfo  {journal} {The Journal of
  Physical Chemistry C}\ }\textbf {\bibinfo {volume} {113}},\ \bibinfo {pages}
  {20998} (\bibinfo {year} {2009}{\natexlab{b}})}\BibitemShut {NoStop}%
\bibitem [{\citenamefont {Brant}\ \emph {et~al.}(2013)\citenamefont {Brant},
  \citenamefont {Giles},\ and\ \citenamefont
  {Halliburton}}]{2013-Brant-JAP-113-5}%
  \BibitemOpen
  \bibfield  {author} {\bibinfo {author} {\bibfnamefont {A.~T.}\ \bibnamefont
  {Brant}}, \bibinfo {author} {\bibfnamefont {N.~C.}\ \bibnamefont {Giles}}, \
  and\ \bibinfo {author} {\bibfnamefont {L.~E.}\ \bibnamefont {Halliburton}},\
  }\href {\doibase 10.1063/1.4790366} {\bibfield  {journal} {\bibinfo
  {journal} {Journal of Applied Physics}\ }\textbf {\bibinfo {volume} {113}},\
  \bibinfo {pages} {053712} (\bibinfo {year} {2013})}\BibitemShut {NoStop}%
\bibitem [{\citenamefont {Koudriachova}\ \emph {et~al.}(2001)\citenamefont
  {Koudriachova}, \citenamefont {Harrison},\ and\ \citenamefont
  {de~Leeuw}}]{2001-Koudriachova-PRL-86-1275}%
  \BibitemOpen
  \bibfield  {author} {\bibinfo {author} {\bibfnamefont {M.~V.}\ \bibnamefont
  {Koudriachova}}, \bibinfo {author} {\bibfnamefont {N.~M.}\ \bibnamefont
  {Harrison}}, \ and\ \bibinfo {author} {\bibfnamefont {S.~W.}\ \bibnamefont
  {de~Leeuw}},\ }\href {\doibase 10.1103/PhysRevLett.86.1275} {\bibfield
  {journal} {\bibinfo  {journal} {Phys. Rev. Lett.}\ }\textbf {\bibinfo
  {volume} {86}},\ \bibinfo {pages} {1275} (\bibinfo {year}
  {2001})}\BibitemShut {NoStop}%
\end{thebibliography}%

\end{document}